\documentclass[iop]{emulateapj}
\usepackage{epsfig}
\usepackage{amsmath}
\usepackage{color}

\begin{document}
\title{Regularization Techniques for PSF--Matching Kernels. I. Choice of Kernel Basis}
\author{
  A.C.~Becker\altaffilmark{1},
  D.~Homrighausen\altaffilmark{2},
  A.J.~Connolly\altaffilmark{1},
  C.R.~Genovese\altaffilmark{2},
  R.~Owen\altaffilmark{1},
  S.J.~Bickerton\altaffilmark{3},
  R.H.~Lupton\altaffilmark{3}
}
\altaffiltext{1}{Astronomy Department, University of Washington, Seattle, WA 98195}
\altaffiltext{2}{Carnegie Mellon University, Department of Statistics. Pittsburgh, PA 15232}
\altaffiltext{3}{Department of Astrophysical Sciences, Princeton University, Princeton, NJ 08544}

\begin{abstract} 

We review current methods for building PSF--matching kernels for the
purposes of image subtraction or coaddition.  Such methods use a
linear decomposition of the kernel on a series of basis functions.
The correct choice of these basis functions is fundamental to the
efficiency and effectiveness of the matching -- the chosen bases
should represent the underlying signal using a reasonably small number
of shapes, and/or have a minimum number of user--adjustable tuning
parameters.
We examine methods whose bases comprise multiple Gauss--Hermite
polynomials, as well as a form free basis composed of
delta--functions.  Kernels derived from delta--functions are unsurprisingly shown to
be more expressive; they are able to take more general shapes and
perform better in situations where sum--of--Gaussian methods are known
to fail.  However, due to its many degrees of freedom (the maximum number
allowed by the kernel size) this basis tends to overfit the problem, and
yields noisy kernels having large variance.
We introduce a new technique to regularize these delta--function
kernel solutions, which bridges the gap between the generality of
delta--function kernels, and the compactness of sum--of--Gaussian
kernels.
Through this regularization we are able to create general
kernel solutions that represent the intrinsic shape of the PSF--matching kernel
with only one degree of freedom, the strength of the regularization
$\lambda$.  The role of $\lambda$ is effectively to exchange variance
in the resulting difference image with variance in the kernel itself.
We examine considerations in choosing the value of $\lambda$,
including statistical risk estimators and the ability of the solution
to predict solutions for adjacent areas.  Both of these suggest
moderate strengths of $\lambda$ between 0.1 and 1.0, although this optimization is
likely dataset dependent.
This model allows for flexible representations of the convolution kernel
that have significant predictive ability, and will prove useful in
implementing robust image subtraction pipelines that must address
hundreds to thousands of images per night.

\end{abstract}

\section{Introduction}

Studies of variability in astronomy typically use image subtraction
techniques in order to characterize the magnitude and type of the
variability.  This practice involves subtracting a prior--epoch
(generally high signal--to--noise) template image from a recent
science image; any flux remaining in their difference may be
attributed to phenomena that have varied in the interim.  This
technique is sensitive to both photometric and astrometric
variability, and can uncover variability of both point--sources
\citep[such as stars or supernovae;
e.g.][]{2008AcA....58...69U,2008AJ....135..348S} and extended--sources
\citep[such as comets or light echoes; e.g.][]{2006PASP..118.1484N}.
Successful application of this technique shows that it is sensitive to
variability at the Poisson noise limit in a variety of astrophysical
conditions
\citep{Alard98,Alard00,2008MNRAS.386L..77B,2010arXiv1004.2166K}, and
in this regard may be considered optimal.

There are several reasons for preferring such an approach over
catalog--based searches.  First, many types of variability are found
in confused regions of the sky, and it may be difficult to deblend the
time--variable signal from the non--temporally--variable surrounding
area.  This is particularly true for supernovae and active galactic
nuclei, which are typically blended with light from their host
galaxies.  However, such confusion is not limited to stationary
objects.  Moving solar system bodies may serendipitously yield false
brightness enhancements in the measurement of a background object if
the impact parameter is small compared to the image's point spread
function (PSF).  For this reason, removal of non--variable objects is
preferred before attempting to characterize variable sources in images.

Image subtraction is also an efficient technique as the vast majority
of pixels in an image do not contain signatures of astrophysical
variability.  Any pixel--level analysis of a difference image will,
therefore, be restricted to those sources that are temporally variable
(as opposed to analyzing all sources within an image).
While many variants of this technique have been published
\citep{Tomaney96,Alard98,2008MNRAS.386L..77B,2009MNRAS.397.2099A}, and
many versions implemented in automated variability--detection
pipelines
\citep{2001MNRAS.327..868B,2005ApJ...634.1103R,2007ApJ...661L..45D,2008PASP..120..449M,2008AcA....58...69U},
there does remain room for improvement in the robustness of the image
subtraction, and in the reduction of subtraction artifacts.  We refer
the reader to \cite{2008mmc..confE...3W} for an in--depth summary on
the practical application of these image subtraction techniques.

\subsection{Image Subtraction}
In image subtraction we assume that we have two images of the same
portion of the sky, taken at different epochs, but in the same filter.
We will call the image that contains the variability of interest the
``science'' image, and the template image to be subtracted the
``reference'' image.  The images will, in general, be astrometrically
misaligned, but this can be resolved by using sinc--based image registration
methods that preserve the noise properties of the original image.
After astrometric alignment, a given astrophysical object will be
represented in the reference image as a sub-array of pixels $R(x,y)$
and in the science image as $S(x,y)$, with the same span in $x$ and
$y$.
Each image will, however, have a different point spread function (PSF),
which is the spatial response of a point source due to the atmosphere,
telescope optics, and instrumental signatures.  PSF--matching of the
images is required before we can subtract one image from the other,
and is the essence of the image subtraction technique.

\section{PSF--Matching}

We typically assume that the reference image is a high
signal--to--noise (S/N) representation of the field, for example an
image coadd made through a mosaicing process, or a single image taken
on a night with particularly good seeing.
A standard assumption \citep[e.g.][]{Alard98,Alard00} is that $S(x,y)$ can be modeled as a convolution of
$R(x,y)$ by a single PSF--matching kernel $K(u,v)$, with an additional
noise component $\epsilon(x,y)$;
\begin{eqnarray}
\label{eqn-conv}
S(x,y) & = & (K \otimes R)(x,y) + \epsilon(x,y).  
\end{eqnarray}
Our goal in this paper is to develop an effective method for
determining $K(u,v)$.

\subsection{Linear Modeling of $K(u,v)$}
As inputs to the PSF--matching technique, we assume images are
astrometrically registered, and background subtracted (while this
latter constraint is not a necessity, it does enable us to restrict
our analysis here to the respective shapes of the PSFs).
To proceed, we make the assumption that $K(u,v)$ may be
modeled as a linear combination of basis functions $K_i(u,v)$, such
that $K(u,v) = \sum_i a_i K_i(u,v)$ \citep{Alard98}.
The basis components do not have to be orthonormal, nor does the basis
need to be complete (indeed, it may be overcomplete).  However, it is desirable
to choose a shape set that compactly describes $K(u,v)$, such that the number of required
terms is small.  

By formulating the kernel decomposition as a linear expansion, we may
recast Equation~\ref{eqn-conv} as the vectorized equation
\begin{eqnarray}
\label{eqn-lin1}
S & = & C a + \epsilon
\end{eqnarray}
where C is the matrix of functions $C_i \equiv (K_i \otimes R)$
evaluated at each pixel.  For any given kernel basis set, the goal is
to find the coefficients $a_i$ associated with each $K_i$.

We proceed using standard linear least squares analysis.  
We assume that the noise is uncorrelated and known; $\epsilon$ is
therefore the product of a diagonal matrix $\Sigma^{1/2}$, which
represents the square root of the known per--pixel variance, and a
zero--mean unit--variance
random variable $Z$.  By reweighting by the inverse square root of
$\Sigma$ (which must exist as covariance matrices are positive
definite and hence invertible) we obtain the modified equation
\begin{eqnarray}
\label{eqn-lin2}
\Sigma^{-1/2} S & = & \Sigma^{-1/2} C a + Z
\end{eqnarray}
which is just another linear model, now with the error term $Z$ having
the identity matrix for the covariance.  This reduces to the weighted
linear least squares equation
\begin{eqnarray}
\label{eqn-lin3}
\tilde{S} & = & \tilde{C} a + Z,  
\end{eqnarray}
with
\begin{eqnarray}
\tilde{S} & \equiv & \Sigma^{-1/2} S, \\ \nonumber
\tilde{C} & \equiv & \Sigma^{-1/2} C. \nonumber
\end{eqnarray}

The normal equations for estimating $a$ are:
\begin{eqnarray}
\label{eqn-lin4}
\tilde{C}^{\top} \tilde{S} & = & \tilde{C}^{\top} \tilde{C} a \\
C^{\top} \Sigma^{-1} S & = & C^{\top} \Sigma^{-1} C a. \nonumber
\end{eqnarray}

This may be cast in the familiar form of $b = M a$ with 
\begin{eqnarray}
\label{eqn-lin5}
b & = & C^{\top} \Sigma^{-1} S \\
M & = & C^{\top} \Sigma^{-1} C. \nonumber
\end{eqnarray}
In discrete pixel coordinates, this corresponds to
\begin{eqnarray}
\label{eqn-mb}
b_{i}  & = & \sum_{x,y} {{C_i(x,y) S(x,y)}\over{\sigma^2(x,y)}}  \\
M_{ij} & = & \sum_{x,y} {{C_i(x,y) C_j(x,y)}\over{\sigma^2(x,y)}} \nonumber
\end{eqnarray}
where $\sigma^2(x,y)$ represents the known variance per pixel.
The creation of the matrices $M_{ij}$ and $b_{i}$ therefore requires a
convolution of the reference image with each basis kernel.

The least--squares estimate for $a$ is $\hat{a} = (\tilde{C}^{\top}
\tilde{C})^{-1} \tilde{C}^{\top} \tilde{S}$.  A difference image is
then constructed as $D(x,y) = S(x,y) - C\hat{a}(x,y)$.  
Because the estimate of $\hat{a}$ is explicitly dependent on both
$S(x,y)$ and $R(x,y)$, the residuals in the difference image may {\it not} necessarily follow a
normal $\mathcal{N}(0, 1)$ distribution\footnote{We use the mean and
variance, not mean and standard deviation, as the two parameters of
Normal distributions.}, with $\sigma^2 \neq 1$ due to this covariance.  The
residuals should however have a flat power spectral density.

\subsection{Invertability of $C^{\top} \Sigma^{-1} C$}
\label{sec-invert}

When a large set of basis functions is used, the matrix $M = C^T
\Sigma^{-1} C$ may be ill--conditioned or even singular.  This can be
quantified by the ``condition number'' of $M$, which we define as the
ratio of the largest to the smallest eigenvalues.  When the condition
number is large, 
inversion of $M$ will be numerically unstable or infeasible.

A common approach when trying to invert an ill--conditioned matrix is
to compute instead a \emph{pseudo--inverse}, or an approximation to
one in which eigenvalues that are numerically small are zeroed out.
As $M$ is symmetric, we can decompose it as $M = V D V^T$ with $V$ an
orthogonal matrix and $D = {\rm diag}(d_1,\ldots,d_n)$ with
eigenvalues $d_1 \geq d_2 \geq \ldots \geq d_n \geq 0$.  We define $D_i =
{\rm diag}(d_1,\ldots,d_i,0,\ldots,0)$ to be a truncation of $D$ where
$d_1 / d_{i+1}$ becomes too large.
Then, we define the pseudo--inverse of $D_i$ as $D_i^{\dagger} = {\rm
diag}(1/d_1,\ldots,1/d_i,0,\ldots,0)$.  Note this allows for the
definition of a pseudo-inverse of $M$ as $M^{\dagger} = V
D_i^{\dagger} V^T$.  Analogous to $D_i$, define $V_i$ to be the same
as the matrix $V$ in the first $i$ columns, and zero elsewhere.
Typically this truncation threshold is defined by the machine
precision of the computation (e.g. for double--valued calculations, $1
/ d_{min} \sim 5 \times 10^{15}$).  However, significantly larger
limits for $d_{min}$ may be used to avoid underconstrained parameters,
such as in Section~\ref{sec-sure}.

\section{Sum--of--Gaussian Bases}
\label{sec-al}

The original PSF--matching bases proposed by \cite{Alard98} and
\cite{Alard00} (referred to here as ``Alard--Lupton'' or AL bases)
used a sum of multiple Gaussians, each modified by a 2--dimensional
polynomial:
\begin{eqnarray}
\label{eqn-al}
K_i(u,v) & = & 
               e^{-(u^2 + v^2)/2 \sigma^2_{n}} ~~ u^{p} v^{q},
\end{eqnarray}
where the index $i$ runs over all permutations of $n, p, q$.  This
basis set effectively uses $n=1 \ldots N$ Gaussian components, each with width
$\sigma_n$, and each modified by a set of Gauss--Hermite polynomials
\citep[e.g.][]{0305-4470-33-8-307} expanded out to order $O_n$ ($0
\leq p + q \leq O_n$).  The total number of basis functions in the set
is $\sum_n (O_{n} + 1) \times (O_{n} + 2) / 2$.

The number $N$ and width $\sigma_n$ of the Gaussians, as well as
spatial order of the polynomials $O_n$, are configurable but are not
fitted parameters in the linear least-squares minimization.  Therefore
these are tuning parameters of the model.  Typically, {\it a-priori}
information such as the widths of the image PSFs is used to choose
these values \citep[e.g.][]{2007AN....328...16I}.  In a representative
implementation \citep{2002SPIE.4836..395S}, three Gaussians are used,
with the narrowest Gaussian expanded out to order 6, the middle to
order 4, and the widest to order 2.  This leads to a total of 49 basis
functions used in the kernel expansion.

\begin{figure*}[t]
  \epsscale{1.10}
  \plotone{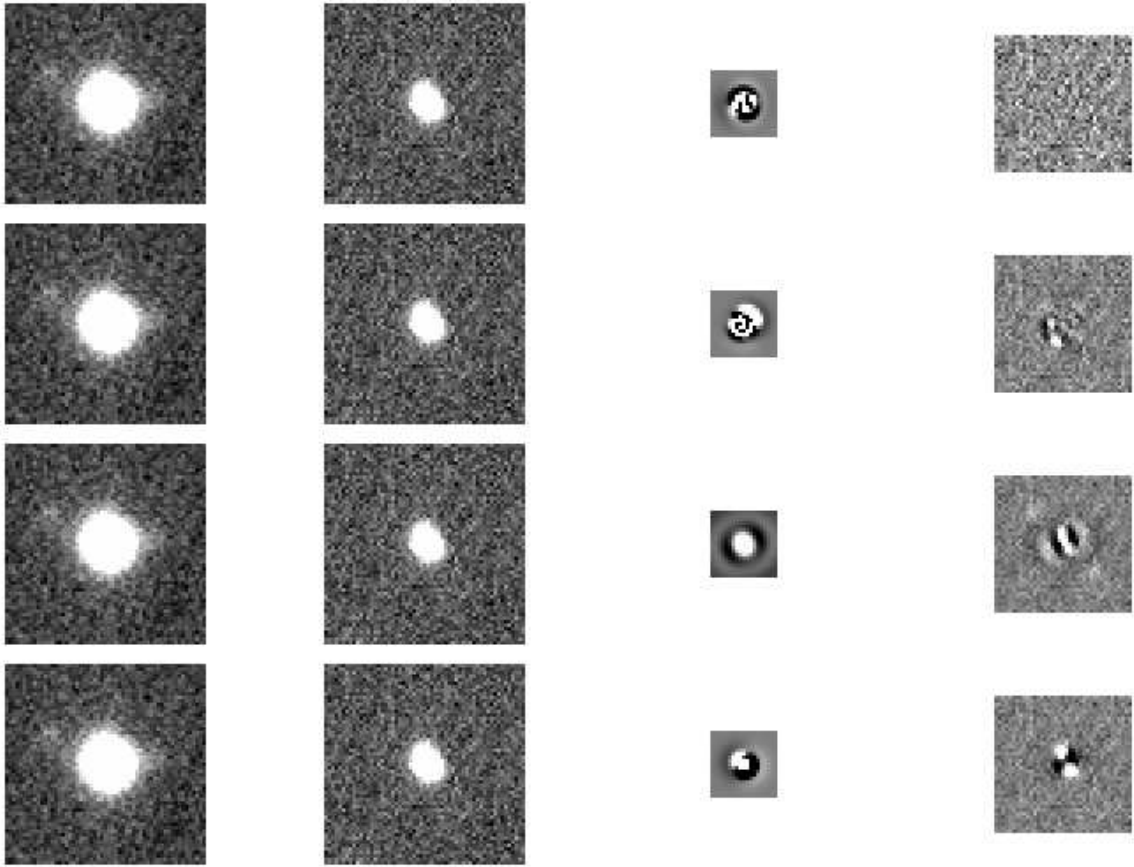}
  \caption{
    Difference imaging results when using a sum--of--Gaussian basis.
    The first column shows the reference image to be convolved
    $R(x,y)$, the second shows the science image $S(x,y)$ the reference
    is matched to, the third column shows the best--fit $19 \times 19$
    pixel PSF matching kernel $K(u,v)$, and the fourth column shows
    the resulting difference image $D(x,y)$.  {\it Row 1}: Results
    when using a basis set with $\sigma_n$ = [0.75, 1.5, 3.0] pixels,
    $O_n$ = [4, 3, 2].  {\it Row 2}: Results when the images are
    mis--registered by 3 pixels in both coordinates, requiring
    significant off--center power in the kernel.  {\it Row 3}: Results
    when the basis Gaussians are too large compared to the actual
    PSF--matching kernel ($\sigma_n$ = [3.0, 5.0] pixels, $O_n$ = [3,
    2]).  {\it Row 4}: Results when the polynomial expansion is not
    carried to high enough order ($\sigma_n$ = [0.75, 1.5, 3.0]
    pixels, $O_n$ = [1, 1, 1]). \\ \\
  }
  \label{fig-AL}
\end{figure*}

The practical application of this algorithm has been very successful,
and it has been used by various time--domain surveys such as MACHO
\citep{1999ApJ...521..602A}, OGLE
\citep{2000AcA....50..421W,2008AcA....58...69U}, MOA
\citep{2001MNRAS.327..868B}, SuperMACHO
\citep{2002SPIE.4836..395S,2005ApJ...634.1103R}, the Deep Lens Survey
\citep{Becker04a}, ESSENCE \citep{2007ApJ...666..674M}, the SDSS--II
Supernova Survey \citep{2008AJ....135..348S}, and most recently
analysis of commissioning data from Pan--STARRS
\citep{2010ApJ...717L..52B}.

The top row of Figure~\ref{fig-AL} shows an instance of successful PSF
matching using this sum--of--Gaussians basis.  The first column
represents a high signal--to--noise image of a star $R(x,y)$ generated
from an image coaddition process applied to data from the
Canada--France--Hawaii Telescope (CFHT).  The second column shows this
same star, aligned with the template image to sub--pixel accuracy, in
a single science image $S(x,y)$.  The star is obviously asymmetric,
potentially due to optical distortions such as focus or astigmatism,
or due to tracking problems during acquisition of the image.  The
PSF--matching kernel thus will need to take the symmetric $R(x,y)$ and
elongate it along a vector oriented approximately $135\deg$ from
horizontal.  The first row, third column shows the best--fit
PSF--matching kernel using $N = 3$ Gaussians with $\sigma_n$ = [0.75,
  1.5, 3.0] pixels, and each modified by Hermite polynomials of order
$O_n$ = [4, 3, 2], respectively.  The total number of terms in the
expansion is 31.  The first row, right column shows the resulting
difference image $D(x,y)$.  The subtraction is obviously very good,
with the remaining pixels described by a $\mathcal{N}(0.01, 1.01)$
distribution.

\subsection{Limitations of the Model}

The intrinsic symmetries of Hermite polynomials (symmetric for even
order, anti--symmetric for odd order) means that the Gauss--Hermite
bases possess a high degree of symmetry about the central pixel.  This
makes it difficult to concentrate the kernel power off--center when using an
incomplete basis expansion.  Such functionality
is necessary when the flux needs to be redistributed on the scale of
the kernel size, such as when there are astrometric misalignments.
While it is possible to compensate for misalignment using kernels
derived from this basis, this requires concentrating
the kernel strength in the high--order terms.  There are practical
limitations to the efficacy of this including the scale and orientation of the required shift, 
and the number of basis terms used.

As a concrete example, the second row in Figure~\ref{fig-AL} shows the
best--fit kernel derived when there is a 3--pixel shift in both the
$x$ and $y$ directions.  The kernel needs to have power in the first
quadrant (upper right) at the scale of 3 pixels.  The image of the
kernel (third column) shows that while it is obviously able to do so,
the matching suffers in the third quadrant, as the difference image
shows obvious residuals.  These pixels result in an unacceptable
$\mathcal{N}(0.01, 1.44)$ distribution; recall we were able to yield
$\sigma^2 = 1.01$ for well--registered images (top row).

Another limitation of the model is that there are a variety of tuning
parameters.  This includes the number of Gaussians in the basis, their
widths, and their spatial orders.  These parameters are typically
chosen using a set of heuristics. If there is a mismatch compared to
the true underlying kernel, this process will fail.  The third row of
Figure~\ref{fig-AL} shows PSF--matching results when the basis
Gaussians are {\it too big} and are unable to reproduce the
small--scale differences in the PSFs.  This yields obvious residuals
in the difference image, which follow a $\mathcal{N}(0.02, 3.03)$
distribution.  The fourth row of Figure~\ref{fig-AL} shows results
when the Gauss--Hermite polynomials are not allowed to vary to high
enough order, also yielding unacceptable $\mathcal{N}(0.02, 2.86)$
residuals in the difference image.

Clearly the results of this process are sensitive to the choice of
several tuning parameters, which makes this difficult to implement
robustly.  In a statistical sense, selection of tuning parameters
(which includes selecting the number of basis functions used) usually
has a much larger effect on performance than does the choice of basis
functions.  A process that results in a reduction in the number of
kernel tuning parameters, while maintaining the quality of the
difference images, would greatly improve the effectiveness of this
method.

\section{Delta--Function Bases}
\label{sec-df}

The most general technique for modeling $K(u,v)$ is to use a ``shape
free'' basis, which consists of a delta function at each kernel pixel
index: $K_{ij}(u,v) = \delta(u-i)\delta(v-j)$.  A kernel of size $19
\times 19$ will then have 361 orthonormal, single--pixel bases.  In
this situation there are {\it no} tuning parameters, which is an
obvious benefit.  
However, in any choice of basis there is a trade off between
flexibility in the forms the fitted function can take, and variability
in the resulting fit (the so--called ``bias--variance'' trade
off). The delta--function basis provides complete flexibility, and as
such can account for features such as arbitrary off-center power
required to compensate for astrometric misregistration
\citep[e.g.][]{2008MNRAS.386L..77B}.  But to avoid gross overfitting,
that flexibility needs to be tempered to keep the variance in check.

\begin{figure*}[t]
  \epsscale{1.10}
  \plotone{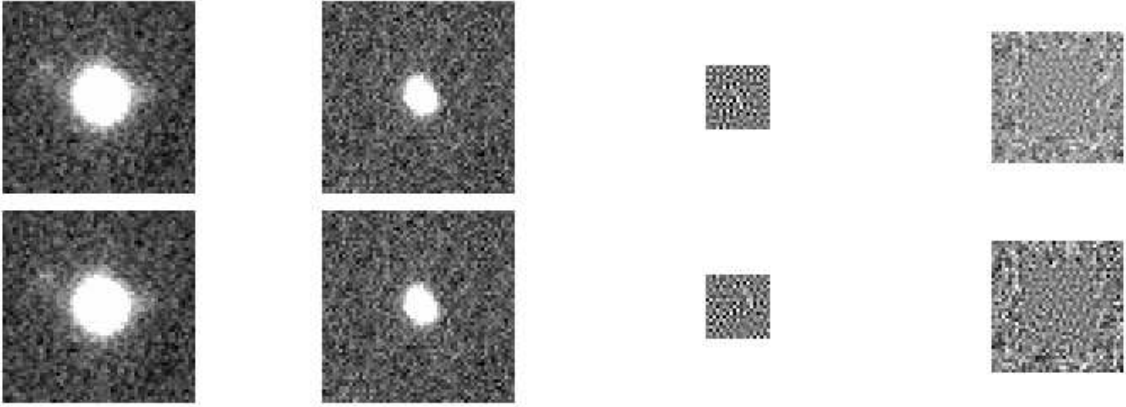}
  \caption{
    Difference imaging results when using a delta--function basis.
    Columns are the same as in Figure~\ref{fig-AL}.  {\it Row 1}:
    Results when using an unregularized delta--function
    basis.  {\it Row 2}: Results when the images are
    mis--registered by 3 pixels in both coordinates. \\ \\ 
  }
  \label{fig-DF}
\end{figure*}

Figure~\ref{fig-DF} shows the results of PSF--matching using such a
basis, using the same objects as in Figure~\ref{fig-AL}.  The top row
demonstrates the results for exactly aligned images, while the bottom
row demonstrates the results for images misaligned by 3 pixels in both
$x$ and $y$.  The difference images are qualitatively similar.
However, the best--fit solutions obviously yield large
variations within the kernels themselves, and do not match expectations of what the 
actual kernel should look like.  The reason for this can be found in the
distribution of pixels residuals in the difference image.  Both images
follow a $\mathcal{N}(0.01, 0.79)$ distribution.  This indicates that
the residuals have lower variance than Gaussian statistics would
suggest.  Indeed, in Figure~\ref{fig-DF} column~4 the residuals appear
smoother than random noise.  This is impossible unless we have
overestimated the variance in our images, or unless the kernels
themselves are removing some fraction of the noise.

The large numbers of basis shapes (361 degrees of freedom vs. 31 for
the sum--of--Gaussians) makes it highly likely that we are
over--fitting the problem.  The kernel thus has the ability to match
both the underlying signal {\it and} the associated noise in the two
images.  So while this technique is optimal for matching pixels in two
images -- where those pixels are a combination of signal and noise --
it is not necessarily optimal for uncovering the true PSF--matching
kernel.

A consequence of this is that the PSF--matching kernel derived for any
given object may not be directly applied to neighboring objects, since
the solution is significantly driven by the local noise properties.
High variance estimators are particularly poor as inputs to
interpolation routines, or to a spatial model of the kernel
$K(u,v,x,y)$, that find the matching kernel at {\it all} locations as
a function of the fitted kernels at particular locations.
Below, we explore how introducing a certain amount of bias into this
estimator can improve its performance.

\section{Delta--Function Bases with Regularization}
\label{sec-dfr}

The delta--function basis can flexibly fit a kernel of any form, but as
we have shown, this flexibility is both its strength and weakness.  As
is, the method significantly overfits, absorbing substantial noise
fluctuations into the fit and thus giving estimated kernels with
excessive variance.  A solution is to introduce some amount of bias
into the fit to reduce the solution variance by a much larger factor
(if "bias" sounds pejorative, note that this is just a kind of
smoothing).  When fitting a smooth function such as $K(u,v,x,y)$, we
prefer fitted kernels for which nearby solutions do not vary too
greatly.  This bias will enable such a fit with vastly reduced
mean--squared error.

Among the various approaches to dealing with overfitting, the most
common are through linear regularization techniques
\citep[e.g. Section 18.5;][]{1992nrca.book.....P}.  Using these, we
may penalize undesirable features of the fit, usually by adding a
penalty term to our optimization criterion.  For instance, when
fitting a smooth function, we want to penalize fits $f$ that are too
rough or irregular.  One way to do this is to add to the least squares
objective a term penalizing the second derivative, $\lambda \cdot \int
|f''(x)|^2 {\rm d}x$.  Here, the scaling factor $\lambda$ is a tuning
parameter that determines the balance between fidelity to the data and
the desired smoothness.  In the case of kernel matching, we may extend
this idea with a two dimensional penalty that approximates $\lambda
\cdot \int\int |\nabla f(x,y)|^2 {\rm d}x {\rm d}y$.

The one--dimensional second derivative of a function $f$ around pixel
$x$ may be approximated using the central finite difference $f''(x)
\approx f(x-1) - 2 f(x) + f(x+1)$.  Since the delta--function bases
have unit height and no intrinsic shape, regularizing the coefficients
$a_i$ is equivalent to regularizing the shape of the resulting kernel
(care must be taken to apply the regularization penalty to only those
pixels that are associated spatially).  In matrix terms, this
one--dimensional regularization may be represented by $R_1 a$, with
\begin{eqnarray}
\label{eqn-r2}
R_1 & = & {
\left( {\begin{array}{rrrrcr}
1        &        -2 &  1~~ &  0 & ~$\ldots$~  & 0        \\
0        &         1 & -2~~ &  1 & ~$\ldots$~  & 0        \\
$\vdots$ &           &   ~~ &    &             & $\vdots$ \\
0        &~$\ldots$  &  0~~ &  1 & -2~         & 1        \\
 \end{array} } \right) }
\end{eqnarray}
which is of dimension $(m - 2) \times m$, where $m$ here is the total
number of pixels in the kernel \footnote{The absolute value of the
kernel's border pixels may also be penalized through the addition of a
row at both the top and bottom of $R_1$.}.  A generalization of this
to two dimensions results in a 5--point stencil that sums the local
derivative along both axes, $f''(x,y) \approx f(x-1,y) + f(x+1,y) +
f(x,y-1) + f(x,y+1) - 4 f(x,y)$, with an associated matrix $R_2$.

The finite calculation of this penalty is implemented through the
matrix equations
\begin{eqnarray}
\label{eqn-r}
|R_2 a|^2 & = a^{\top} R_2^{\top} R_2 a
\end{eqnarray}
where $a$ represents the amplitude of each delta function, and $R_2$
encapsulates the coefficients that approximate the local $2^{nd}$
derivative of the resulting kernel.
We define the matrix $H \equiv R_2^{\top} R_2$, which makes the second
derivative penalty $a^{\top} H a$.  This matrix is used to regularize
the normal equations (Equation~\ref{eqn-lin4}) with strength $\lambda$
\begin{eqnarray}
\label{eqn-reg}
C^{\top} \Sigma^{-1} S & = & (C^{\top} \Sigma^{-1} C + \lambda H) a \\
b & = & M_{\lambda} a \nonumber.
\end{eqnarray}
Note the similarity to Equation~\ref{eqn-lin3}, with the only
difference being $M_{\lambda} = M + \lambda H$.  Here $\lambda$
represents the strength of the regularization penalty, and is the sole
tuning parameter in this model.

\begin{figure*}[t]
  \epsscale{1.10}
  \plotone{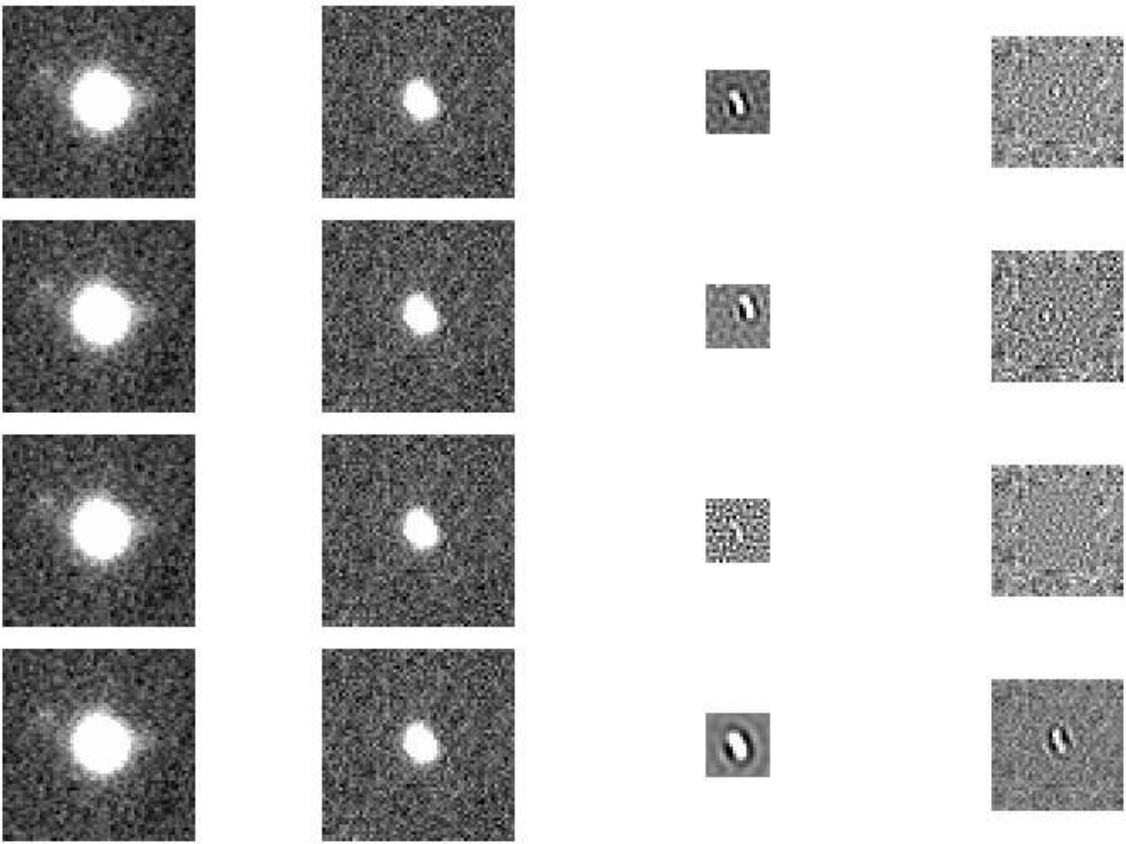}
  \caption{
    Difference imaging results when using a regularized
    delta--function basis.  Columns are the same as in
    Figure~\ref{fig-AL}.  {\it Row 1}: Results when using a
    regularized delta--function basis with $\lambda = 1.0$.  {\it Row
    2}: Results when the images are mis--registered by 3
    pixels in both coordinates, $\lambda = 1.0$.  {\it Row 3}: Results
    using ``weak'' regularization of the kernel, with $\lambda =
    0.01$.  {\it Row 4}: Results using ``strong'' regularization of
    the kernel, with $\lambda = 100$. \\ \\ 
  }
  \label{fig-DFr}
\end{figure*}

Figure~\ref{fig-DFr} shows results for the same set of objects
displayed in Figure~\ref{fig-AL} and Figure~\ref{fig-DF}, but using
regularization of the delta--function basis set.  The top row shows
the results for aligned images, and $\lambda = 1$.  Note that the
kernel looks very much as anticipated, being compact and having a
shape aligned approximately $135 \deg$ from horizontal.  Residuals in
the difference image follow a $\mathcal{N}(0.01, 0.94)$ distribution.
The second row shows the results when the images are misaligned by 3
pixels in $x$ and $y$.  The kernel merely appears shifted by the same
amount compared to the aligned images, and the difference image
follows a quantitatively similar $\mathcal{N}(0.01, 0.96)$
distribution.  This effectively demonstrates that this method can
reproduce kernels with off--center power.  The third row shows the
results with $\lambda = 0.01$; the shape of the PSF--matching
component of the kernel is just barely discernible above its noise,
suggesting the regularization is too weak.  The difference image is,
however, acceptable ($\mathcal{N}(0.01, 0.81)$).  The fourth row shows
the results with $\lambda = 100$.  The kernel is far smoother than in
previous runs.  However, this appears to be at the expense of
residuals in the difference image, which follow a $\mathcal{N}(0.01,
1.35)$ distribution.  This suggests that too much weight has been
given the smoothness of the kernel compared to the residuals in the
difference image, indicating that the regularization is too strong.
The general trend is that with increasing lambda, the variance in the
difference image increases.  The noise properties of the difference
image evolve from being too smooth, to approximately white in
spectrum, to having residual features at a similar scale as the
kernel.

Overall, this technique appears very effective.  We are able to create
general, compact kernels that represent the underlying shape of the
PSF--matching kernel with only one tuning parameter, the strength of
the regularization $\lambda$.
The role of $\lambda$ is effectively to exchange variance in the
resulting difference image with variance in the kernel itself.  By
increasing the value of $\lambda$, we are able to smooth the kernel
while increasing the variance in the difference image.
We explore various methods to establish the optimal value of $\lambda$
below.

\subsection{Choice of Tuning Parameter $\lambda$}
\label{sec-chooselam}

Choosing a good tuning parameter is essential for good performance of
a regularization method.  If $\lambda$ is too high, the fit will be
too smooth (high bias, low variance); if $\lambda$ is too low, the fit
will be too rough (low bias, high variance).  The goal of data--driven
methods for choosing tuning parameters is to find the sweet spot in
the bias--variance trade off.  While choosing a good value for
$\lambda$ is a hard statistical problem, there are a variety of
methods that have proven successful in practice.  These methods
construct a statistical estimate of mean--squared error and choose
$\lambda$ to minimize it.  For instance in cross--validation
\citep[reviewed in][]{Kohavi95astudy}, the data set is broken into
pieces, and each piece is left out in turn during the fit.  The
(prediction) mean--squared error is derived from the average squared
error of the fits in predicting the part of the data that was left
out.  Another approach, called empirical risk estimation \citep{SURE},
uses the data itself to compute an (unbiased) estimate of original
fit's mean--squared error and chooses $\lambda$ to minimize it.  The
theoretical justification for these methods is that, when properly
done and with sufficiently large data sets, the chosen $\lambda$ is
close to the value that minimizes the corresponding mean--squared
error function.

A second tuning consideration is that frequently a set of fitted kernels
will be used to constrain a spatial model $K(u,v,x,y)$ that will be
applied to {\it all} pixels in an image.  Therefore we must give a
large weight to our ability to interpolate between the ensemble of
kernel realizations used to constrain $K(u,v,x,y)$.  One metric for
this is to examine the predictive power of a kernel derived from one
object, and applied to a neighboring object.  At small separations,
the quality of each difference image should be similar, indicating
that the initial solution was not significantly driven by the local
noise properties.  

We explore the practical application of these ideas below using
several sets of CCD images from the Canada--France--Hawaii Telescope
plus Megacam imager, calibrated using the ELIXIR pipeline of
\cite{2004PASP..116..449M}.  The template image used is the median of
several images into a single high S/N representation of the field.
The variance per pixel is determined from the image pixel values
divided by the gain.

\subsubsection{Empirical Risk Estimation}
\label{sec-sure}

\begin{figure*}[t]
  \epsscale{1.10}
  \plotone{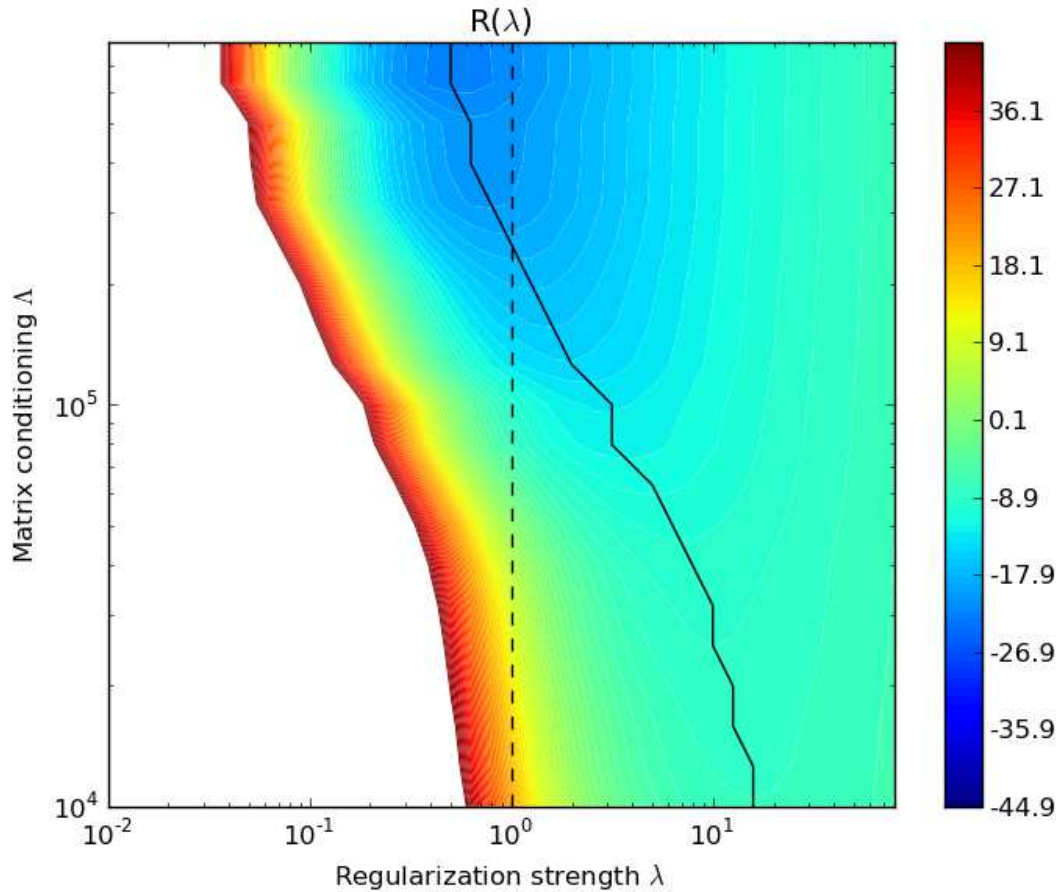}
  \caption{
    Values of the empirical risk $R(\lambda)$, as defined in
    Equation~\ref{eqn-sure}, for different values of the matrix
    conditioning parameter $\Lambda$, and the regularization strength
    $\lambda$.  At all $\Lambda$, we determine the minimum values of
    $R(\lambda)$, which are connected by the {\it solid} black line.
    The {\it dotted} vertical line represents the fiducial value of
    $\lambda = 1$.  The global minimum of $R(\lambda)$ is realized
    with minimal matrix conditioning, and at a value of $\lambda =
    0.5$. \\ \\
  }
  \label{fig-risk}
\end{figure*}

We first construct a loss function that represents the sum of squared
differences between the true (unknown) kernel coefficients $a$ and
$\hat a_\lambda$, which is our estimate of the kernel coefficients
when the tuning parameter is set to the value $\lambda$:
\begin{eqnarray}
L(a, \hat a_\lambda) & = & (\hat a_\lambda - a)^{\top} (\hat a_\lambda - a).
\end{eqnarray}
The expectation value of $L(a, \hat a_\lambda)$ is the statistical
risk we will minimize through our choice of $\lambda$ \footnote{It
  should be noted that other risk estimators may be constructed,
  e.g. ones that maximize the quality of the full difference image.}.
When $M$ is well--conditioned, we can construct an unbiased estimator
of the true risk $\langle L(a, \hat a_\lambda) \rangle$ as \cite[Section 2,][]{SURE}
\begin{eqnarray}
\label{eqn-sure}
R(\lambda) & = & ||\hat a_\lambda||_2^2 + \\ \nonumber
           &   &  2\left( {\rm trace} \bigl( M_{\lambda} \bigr) - \hat a_\lambda^{\top} M^{-1} b \right).
\end{eqnarray}

We note that this
estimator of risk does not require tuning parameters.  If we let $\hat
\lambda$ be the minimizer of $R$, then we choose $\hat a_{\hat
\lambda}$ as the estimate of $a$.

For the circumstance that $M$ is ill-conditioned, we present an
adjustment to $R$ from Equation~\ref{eqn-sure}.  Following the
notation from Section~\ref{sec-invert}, for any $i$ define $\Pi = V_i
V_i^{\top}$.  This corresponds to $\Pi$ being a projection matrix onto
the space of the eigenvectors of $M$ that correspond to its $i$
largest eigenvalues.  Note that $i$ is now an additional tuning
parameter, corresponding to choice of condition number (denoted by symbol $\Lambda$) 
for matrix $M$ (Section~\ref{sec-invert}).
A biased estimate of the statistical risk is then:
\begin{eqnarray}
\tilde R(\lambda) & = & ||\Pi \hat a_\lambda||_2^2 + \\ \nonumber
           &   &  2\left( {\rm trace} \bigl( \Pi M_{\lambda} \bigr) - 
                         \hat a_\lambda^{\top} M^{\dagger} b \right).
\end{eqnarray}
While introducing bias into the estimator of statistical risk seems
bad, it can be necessary in situations where $M$ is ill--conditioned.
Small eigenvalues of $M$ corresponds to there being very little
information along the associated eigenvectors.  By zeroing out these
eigenvalues we are effectively saying we cannot reliably estimate with
this little amount of information.  Hence, we concentrate on getting
the estimation correct on the eigenvectors with larger eigenvalues.

For each object detected in the CFHT images, and for given values of
condition number $\Lambda$ ranging from $4 \leq \log(\Lambda) < 6$, we evaluate
$R(\lambda)$ at values of $-2 \leq \log(\lambda) < 2$.
Figure~\ref{fig-risk} shows a typical outcome of this analysis for a
single object.  Along the y--axis we show the associated value of the
conditioning parameter $\Lambda$, and along the x--axis the value of
$\lambda$ at which $R(\lambda)$ is evaluated.  The {\it solid} line
shows the minimum value of $R(\lambda)$ for each $\Lambda$.

We note that as we decrease the acceptable matrix condition number,
thereby truncating more eigenvalues from the matrix pseudoinverse, the
optimum value of $\lambda$ increases.  For matrices with effectively
no conditioning (large $\Lambda$), the optimal value of $\lambda$ is
near $\lambda = 0.5$.  This is in fact the global minimum of the risk.
A similar result is obtained by looking at all objects within an image
and summing their cumulative risk surfaces.  We regard $\lambda = 0.5$
as the value preferred by the empirical risk estimation technique,
with a range of nearly--equivalent risk between $0.3 < \lambda < 1.0$.

\subsubsection{Predictive Ability}
\label{sec-predict}

In most PSF--matching implementations, several dozen objects across a
pair of registered images are used to create individual $K(u,v)$;
ideally these should evenly sample the spatial extent of the images.
Due to spatial variation in the PSFs of the images, caused by optical
aberrations or bulk atmospheric effects, the single kernel that
PSF--matches {\it all} objects in an image must itself vary spatially.
In this case each of the kernels $K(u,v)$ are used to build spatially
varying PSF--matching kernel $K(u,v,x,y)$.  This is typically
implemented as spatial variation on the kernel coefficients
$K(u,v,x,y) = \sum_i a_i(x,y) K_i(u,v)$.  Therefore an additional
consideration in the choice of $\lambda$ is the ability to build
spatial models for the coefficients $a(x,y)$.

To quantify this, we examine the predictive ability of the kernel
solution $\hat a_\lambda$.  In all CFHT images, we identify object
pairs separated by more than 5 pixels but less than 50, a range of
separations where we expect the intrinsic spatial variation of the
underlying kernel to be minimal.  The kernel derived for each object
in a pair is applied to its complement, and the quality of each
difference image assessed.  For components A and B of each object
pair, this yields difference image $D_{AA}$ which is the difference
image of object A with kernel A, $D_{AB}$ which is the difference
image of object A with kernel B, and analogous images $D_{BA}$ and
$D_{BB}$.  We assess the quality of each difference image using the
width of the pixel distribution normalized by the noise, defined as
e.g. $\sigma_{AA}$, within the central $7 \times 7$ pixels of the
difference image.  While we don't expect this distribution to have a
width of exactly 1.0 due to covariance between the solution and the
input images, we do desire that the quality of $D_{AB}$ and $D_{BA}$
should not be significantly worse than that of $D_{AA}$ and $D_{BB}$.

\begin{figure*}[t]
  \epsscale{1.10}
  \plotone{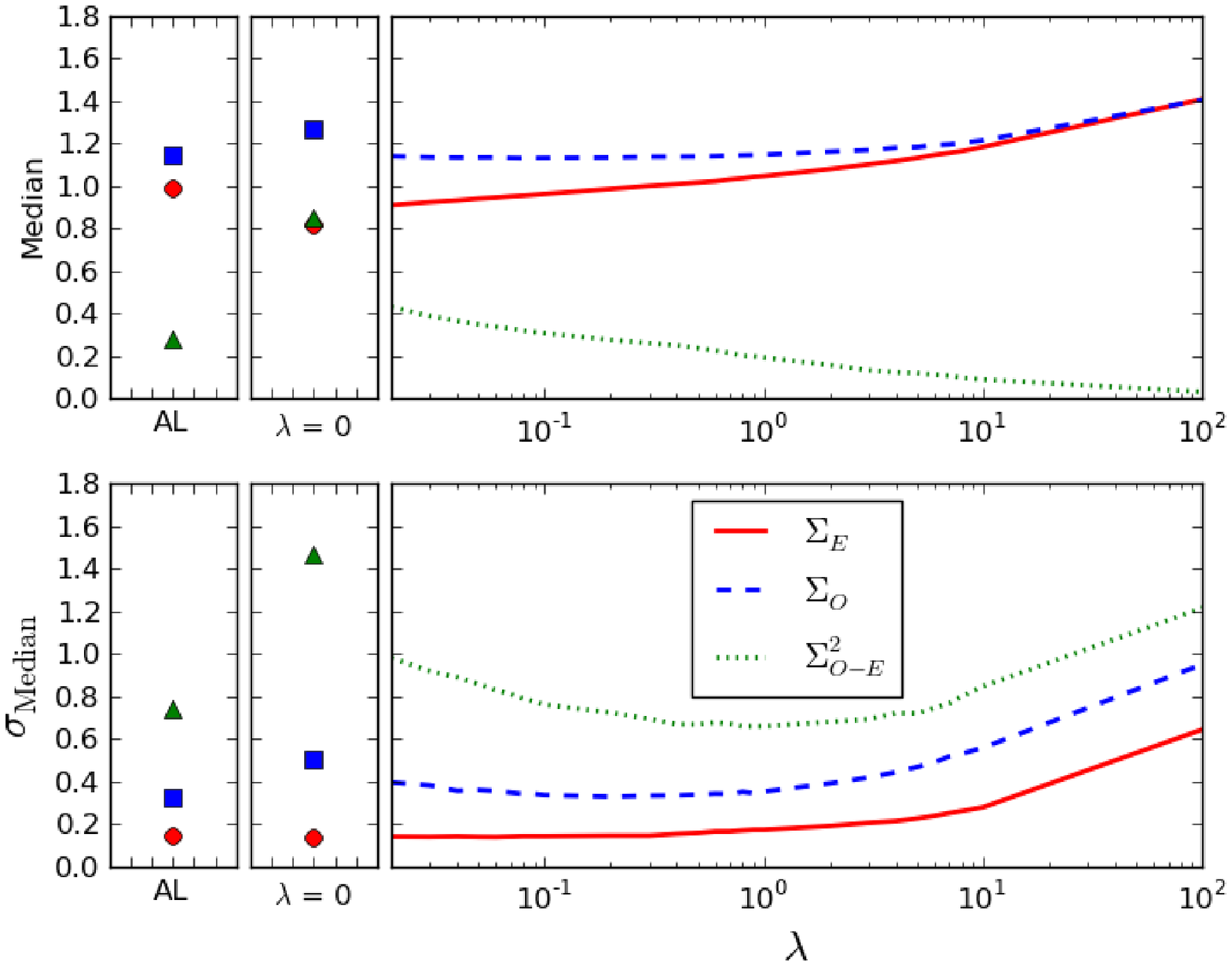}
  \caption{
    Median statistics assessing the predictive ability of different
    kernel bases.  The top panel shows the median values of statistics
    $\Sigma_E$ (red circle and solid line), $\Sigma_O$ (blue square
    and dashed line), and $\Sigma^2_{O-E}$ (green triangle and dotted
    line) for ``Alard--Lupton'' (AL) bases, for delta--function bases
    with $\lambda = 0$, and then for a range of $-2 < \log(\lambda) <
    2$.  All statistics are defined in Section~\ref{sec-predict}.  The
    bottom panel shows the standard deviation of the distribution, defined
    as $74\%$ of the interquartile range. \\ \\
  }
  \label{fig-stats}
\end{figure*}

We aggregate the ``even'' statistics $\sigma_{AA}$ and $\sigma_{BB}$
into distribution $\Sigma_E$, and the ``odd'' statistics
($\sigma_{AB}$, $\sigma_{BA}$) into $\Sigma_O$.  We further examine
the distribution of $\Sigma^2_{O-E}$, which is created from all
measurements of $\sigma^2_{AB} - \sigma^2_{AA}$ and $\sigma^2_{BA} -
\sigma^2_{BB}$.  This statistic reflects the deterioration in an
object's difference image when using a counterpart's kernel, compared
to the optimal kernel derived for that object.

We plot the distributions of these values in Figure~\ref{fig-stats}.
The {\it top} panel provides the median values of these distributions
for the sum--of--Gaussian (AL) basis ({\it left}), for the
unregularized delta function basis ($\lambda = 0$; {\it center}), and
for delta--function regularization strengths of $-2 < \log(\lambda) <
2$ ({\it right}).  The {\it bottom} panel plots the effective standard
deviation of the distribution, defined as $74\%$ of the interquartile
range.

The lowest median residual variance $\Sigma_E$ comes from difference
images made using an unregularized $\lambda = 0$ basis, the reasons
for which we have examined in detail in Section~\ref{sec-df}.
However, as expected the predictive ability of this basis is by far
the worst, having the highest median $\Sigma^2_{O-E}$, as well as
large variance within this distribution.  As we ramp up the
regularization strength, the predictive ability of the kernels
increases (low $\Sigma^2_{O-E}$), but at the expense of the quality of
the difference image itself (large $\Sigma_E$).

To find an acceptable medium between these two considerations, we will
use the results from the sum--of--Gaussian (AL) basis as a benchmark,
since it has been shown to produce effective spatial models
(Section~\ref{sec-al}).  For the AL basis, the median values of
$\Sigma_E$, $\Sigma_O$, and $\Sigma^2_{O-E}$ are 0.99, 1.14, and 0.28,
respectively.  Similar results are obtained with delta--function
regularization strengths of $\lambda \approx 0.2, 0.7$, and $0.2$.
For AL the $\sigma_{\rm median}$ values of $\Sigma_E$, $\Sigma_O$, and
$\Sigma^2_{O-E}$ are 0.14, 0.33, and 0.74, respectively.  These are
matched (or bested) in the regularized basis for $\lambda \leq 0.2$,
$\lambda = 0.2$, and $0.2 \leq \lambda \leq 6$, respectively.  

In summary, using delta--function regularization strengths of $\lambda
\approx 0.2$, we are able to achieve difference images with a similar
quality to those yielded by the sum--of--Gaussian AL
basis (using $\Sigma_E$ as our metric).  These models have similar
predictive ability when applied to neighboring objects (quantified
using $\Sigma_O$ and $\Sigma^2_{O-E}$), making them useful for
full--image spatial modeling.  Finally, they are seen to be generally
applicable, having a small variance in the above statistics when
evaluated over several hundreds of object pairs.

\section{Conclusions}

We've examined here the choice of basis set on the quality of
PSF--matching kernels and their resulting difference images.  These
include the traditional sum--of--Gaussian (``Alard--Lupton'') basis
and a digital basis based upon delta--functions.  We find that while
the delta--function kernels are the most expressive, they are also the
least compact in terms of localization of power within the kernel.
Having one basis component per pixel in the kernel, they tend to
overfit the data and are more sensitive to the noise in the images
instead of the intrinsic PSF--matching signal.

We introduce a new technique of linear regularization to impose
smoothness on these delta--function kernels, at the expense of
slightly higher noise in the difference images.  These regularized
shapes are shown to be flexible, and yield solutions with sufficient
predictive power to prove useful for spatial interpolation.
We outline two methods to determine the strength of this
regularization that minimize the statistical risk of the kernel
estimate, and that examine the predictive ability of the derived
kernels.  Both methods suggest values of $\lambda$ that are between
0.1 and 1.0.  

Given the large range of image qualities used in image subtraction
pipelines compared to the small number of images used in the analysis
here, we caution that these estimates may not be applicable under all
conditions and should really be estimated on a dataset--by--dataset
basis.  The optimal value of $\lambda$ will be a function of the S/N
in the template and science images, which should affect the level of
kernel smoothing needed, and of the respective seeings in the input
images, which may impact the suitability of our finite--difference
smoothness approximation.

While this implementation appears successful and practical, there are
various improvements we might consider in our regularization efforts.
This includes changing the scale over which the regularization stencil
is calculated based upon the seeing in the images; currently this is
being done in pixel--based coordinates, and not adjusted depending on
the full--width at half--maximum of the input PSFs.  We also plan to
examine additional metrics to determine the optimal value of
$\lambda$, including the power spectrum of noise in the resulting
difference image, which should be flat.  Ultimately, the overall
quality of the {\it entire} difference image is the optimal metric to
use in assessing choice of basis; we will be expanding our analysis to
include full--image metrics and spatial modeling of the kernel.

Finally, the wealth of statistical techniques to efficiently choose
basis shapes has not been exhausted.  Other potential methods include
the use of overcomplete bases, where the choice of the correct subset
of components to use is made though through basis pursuit
\citep{Chen98atomicdecomposition}, as well as the process of ``basis
shrinkage'' through the use of multi--scale wavelets
\citep{Donoho94idealspatial,Donoho95adaptingto}.  In all
considerations, it is an advantage to yield solutions that, as an
ensemble, have a low dimensionality so that spatial modeling is
efficient and spatial degrees of freedom are not being used to
compensate for an inefficient choice of basis.  However, for any given
basis set the choice of regularization (none at all or using a fixed
set of functions) is likely to be the proper place for optimization.

\acknowledgements This material is based, in part, upon work supported
by the National Science Foundation under Grant Number AST-0709394.



\end{document}